\newcommand{\todo}[2][]{\textcolor{blue}{
\ifx&#1&%
  TODO:
\else
  TODO(#1):
\fi
#2}}

\newcommand{\review}[2][]{\textcolor{red}{
\ifx&#1&%
  REVIEW:
\else
  REVIEW(#1):
\fi
#2}}

\documentclass[10pt,conference]{IEEEtran}

\usepackage[colorlinks,urlcolor=blue,linkcolor=blue,citecolor=blue]{hyperref}
\usepackage{adjustbox}
\usepackage{color,array}
\usepackage{amsfonts}
\usepackage{graphicx}
\usepackage{amsmath} 
\usepackage{pifont}
\usepackage{xcolor}
\usepackage{dsfont}
\usepackage{multirow}
\usepackage{booktabs}
\definecolor{myblue}{RGB}{28,226,255}
\usepackage[numbers,sort&compress]{natbib}

\newcommand{\firstletter}[2][]{\textcolor{myblue}{
\ifx&#1&%

\else

\fi
#2}}

\begin{document}

\title{Industrial LLM-based Code Optimization under Regulation: A Mixture-of-Agents Approach} 


\author{
\IEEEauthorblockN{
    Mari Ashiga\IEEEauthorrefmark{4}\IEEEauthorrefmark{2},
    Vardan Voskanyan\IEEEauthorrefmark{2}, 
    Fateme Dinmohammadi\IEEEauthorrefmark{4}, 
    Jingzhi Gong\IEEEauthorrefmark{5}\IEEEauthorrefmark{2}, 
    Paul Brookes\IEEEauthorrefmark{2}, \\
    Matthew Truscott\IEEEauthorrefmark{2}, 
    Rafail Giavrimis\IEEEauthorrefmark{3}\IEEEauthorrefmark{2},  
    Mike Basios\IEEEauthorrefmark{2}, 
    Leslie Kanthan\IEEEauthorrefmark{2}, and    
    Wei Jie\IEEEauthorrefmark{4}$^*$
}
\IEEEauthorblockA{\IEEEauthorrefmark{4}University of West London, UK, \IEEEauthorrefmark{5}University of Leeds, UK, \IEEEauthorrefmark{2}TurinTech AI, London, UK, \IEEEauthorrefmark{3}University of Surrey, UK}

\IEEEauthorblockA{
Emails: \{mari.ashiga, fateme.dinmohammadi, wei.jie\}@uwl.ac.uk, j.gong@leeds.ac.uk, \\
\{vardan, paul, matthew.truscott, rafail, mike, leslie\}@turintech.ai
}
\thanks{$^*$Corresponding author: Wei Jie (wei.jie@uwl.ac.uk).}
}

\markboth{Journal of IEEE Transactions on Artificial Intelligence, Vol. 00, No. 0, Month 2020}
{First A. Author \MakeLowercase{\textit{et al.}}: Bare Demo of IEEEtai.cls for IEEE Journals of IEEE Transactions on Artificial Intelligence}

\maketitle

\begin{abstract}
Recent advancements in Large Language Models (LLMs) for code optimization have enabled industrial platforms to automate software performance engineering at unprecedented scale and speed. Yet, organizations in regulated industries face strict constraints on which LLMs they can use---many cannot utilize commercial models due to data privacy regulations and compliance requirements, creating a significant challenge for achieving high-quality code optimization while maintaining cost-effectiveness.
We address this by implementing a Mixture-of-Agents (MoA) approach that directly synthesizes code from multiple specialized LLMs, comparing it against TurinTech AI's vanilla Genetic Algorithm (GA)-based ensemble system and individual LLM optimizers using real-world industrial codebases.
Our key contributions include: (1) First MoA application to industrial code optimization using real-world codebases; (2) Empirical evidence that MoA excels with open-source models, achieving 14.3\% to 22.2\% cost savings and 28.6\% to 32.2\% faster optimization times for regulated environments; (3) Deployment guidelines demonstrating GA's advantage with commercial models while both ensembles outperform individual LLMs; and (4) Real-world validation across 50 code snippets and seven LLM combinations, generating over 8,700 variants, addresses gaps in industrial LLM ensemble evaluation.
This provides actionable guidance for organizations balancing regulatory compliance with optimization performance in production environments.

\end{abstract}



\section{Introduction}
\label{sec:intro}
Code optimization remains a critical challenge in industrial software development, where developers must balance performance improvements with development velocity and code maintainability \cite{artemis_2025}. In production environments, manual code optimization is time-consuming and requires deep expertise across multiple domains, from algorithmic efficiency to security best practices. Organizations in regulated industries face additional constraints, as data privacy regulations and compliance requirements often restrict the use of commercial AI models for automating these processes \cite{song2025modelgovernance}.

Recent advancements in Large Language Models (LLMs) demonstrate significant potential for optimizing code \cite{survey_jing2025}. However, industrial codebases present unique challenges including complex legacy systems, diverse programming paradigms, and regulatory compliance constraints that single models often struggle to address. Research has shown that ensemble approaches combining multiple LLMs can overcome these limitations by leveraging complementary strengths~\cite{llm_ensemble_ashiga2025}. Techniques such as Mixture-of-Agents (MoA)~\cite{mixtureofagents_wang2024}, LLM cascades~\cite{llmcascades_yue2024}, and evolutionary model merge~\cite{GAMoEs_akiba2024} that uses GA approach \cite{GA_holland1975} have shown promising results by combining different models' capabilities to handle the multifaceted nature of real-world code optimization, from legacy system compatibility to compliance requirements.

However, these ensemble approaches have predominantly been evaluated on text generation tasks such as instruction-following \cite{mixtureofagents_wang2024, dare_yu2024, llmblender_jiang2023} and reasoning \cite{GAMoEs_akiba2024, knowledgefusionwan2024,llmcascades_yue2024}. Even studies that extend these methods to code-related tasks typically focus on competitive programming problems with public datasets \cite{mixtralexperts_jiang2024, ties_merge_yadav2023, fusechat_wan2024}. The effectiveness of recent ensemble techniques like MoA remains largely unexplored in the context of optimizing real-world codebases, particularly where regulatory constraints limit model selection and practical requirements such as performance, maintainability, and resource efficiency significantly influence optimization strategies.

Another challenge involves validation of industrial application. Recent surveys of language models for code optimization reveal a significant gap in real-world validation, with 68\% of studies (36 out of 53 reviewed) evaluated only on competitive programming problems~\cite{perfrl_duan2025, effiLearner_huang2025}, synthetic programs~\cite{llmaided_hong2024,performancealigned_nichols2024}, or optimization algorithms~\cite{mathdiscovery_romera2024, evolcode_hemberg2024} rather than sophisticated real-life software projects~\cite{survey_jing2025}. This preference for non-real-world datasets, while offering rich availability and reproducibility, overlooks critical real-world validations necessary for demonstrating the robustness and applicability of optimization methods in complex industrial scenarios.



To address these challenges, we apply and evaluate a state-of-the-art LLM ensemble approach \cite{mixtureofagents_wang2024} for industrial code optimization on real-world projects. In summary, our key contributions are:
\begin{itemize}
    \item We implement a Mixture-of-Agents (MoA) approach for code optimization that synthesizes outputs from multiple specialized LLMs to improve performance across real-world industrial projects.
    \item We conduct an extensive empirical study comparing MoA with the company's vanilla GA-based ensemble system and standalone LLM optimizers, involving 50 industrial code snippets and over 8,700 generated variants.
    \item We analyze how different model combinations (open-source vs. commercial) impact optimization quality, cost, and compliance, identifying use-case scenarios where MoA excels under regulatory constraints.
    \item We provide practical guidance on deploying ensemble-based optimization in industry, highlighting trade-offs between optimization effectiveness, resource efficiency, and model accessibility.
\end{itemize}


\section{Background and Related Work}
\label{sec:relatedwork}
\textbf{LLM Ensemble Approaches.} Recent advances in ensemble learning for Large Language Models have demonstrated significant improvements in both text \cite{taskarithmetic,GAMoEs_akiba2024,dare_yu2024, knowledgefusionwan2024, switchtransformer_fedus2022, rewardmodelensembles_coste2024, llmblender_jiang2023, mixtureofagents_wang2024,rewardmodelensembleRlhf_zhang2024, routoo_mohammadshahi2024, llmcascades_yue2024, frugalgpt_chen2023} and code generation tasks \cite{mixtralexperts_jiang2024, dare_yu2024, ties_merge_yadav2023}. These approaches address the limitations of individual LLMs, including output inconsistencies and inherent biases, while enabling organizations to leverage diverse model capabilities \cite{llm_ensemble_ashiga2025}. However, most existing ensemble approaches have not specifically addressed the constraints faced by organizations in regulated environments where commercial model usage may be restricted.

\textbf{Mixture-of-Agents Methodology.} MoA \cite{mixtureofagents_wang2024} is one of the state-of-the-art LLM ensemble approaches. It constructs layered architectures where each layer comprises multiple LLM agents that leverage outputs from previous layers as auxiliary information. This methodology has achieved state-of-the-art performance on instruction-following benchmarks, surpassing GPT-4o-mini on AlpacaEval 2.0 \cite{alpacaeval2023} and MT-Bench \cite{mt-bench_zheng2023}. We selected MoA for our implementation due to its potential applicability in code optimization, direct API endpoint integration, and model-agnostic design that makes industrial deployment smooth in our Artemis platform. Importantly, MoA's flexible architecture allows for effective combination of diverse models, including open-source alternatives that can be deployed in environments with regulatory constraints.

\textbf{Artemis Platform.} Our work builds upon the established Artemis AI framework~\cite{artemis_2025}, an industrial platform that distinguishes itself through its multi-stage process: extracting target code snippets, independent optimization by multiple LLMs, and search-based selection of optimal solutions. 
Unlike single-model approaches, Artemis leverages multi-LLM collaboration while preserving code readability and reliability—critical for industrial deployment. 
The platform has demonstrated substantial performance improvements, achieving up to 52\% execution time reduction on real-world projects. In this study, we explore the applicability of the MoA approach to Artemis, comparing it with the company's vanilla GA-based ensemble system to evaluate enhancement potential, particularly in scenarios where regulatory requirements limit model selection.

\section{Mixture-of-Agents Approach} 
Our MoA approach implements a multi-layered ensemble architecture that transforms code optimization from single-model generation into collaborative multi-agent synthesis, leveraging complementary LLM strengths while mitigating individual model limitations.

\subsection{Feedforward Workflow Architecture}
\begin{figure}[!t]
  \centering
  \includegraphics[width=0.5\textwidth]{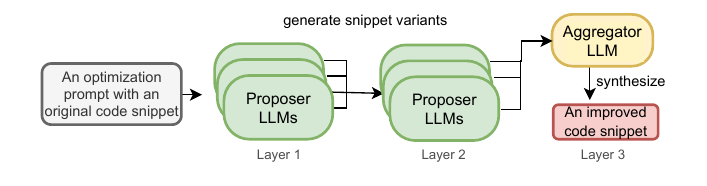}
  \caption{Feedforward structure of MoA with three layers, applied for code optimization. We compare MoA with the company's vanilla GA-based ensemble system and individual LLMs.}
  \label{fig:moa}
\end{figure}

The MoA framework operates as a feedforward pipeline comprising multiple layers of LLM agents. As illustrated in Figure~\ref{fig:moa}, the architecture uses three layers by default, with layer count serving as a tunable hyperparameter.

The process begins with an optimization prompt containing the original code snippet. In the first layer, multiple proposer agents operate in parallel to generate distinct optimization candidates, creating a diverse pool of initial suggestions. In subsequent layers, each agent receives all generated candidates from previous layers as additional context, enabling iterative refinement where agents build upon prior suggestions and integrate strengths from different variants.

The final layer employs a specialized aggregator LLM that synthesizes refined outputs into a single optimized version using a dedicated synthesis prompt. This aggregation combines, reconciles, and enhances the most effective ideas from preceding generations to produce the final output.

\subsection{Agent Specialization and Synthesis Strategy}
Different LLMs exhibit distinct optimization characteristics based on their training data and architecture. Commercial LLMs typically demonstrate superior reasoning capabilities but incur higher costs, while open-source alternatives provide cost-effective solutions—particularly relevant for organizations in regulated environments with restricted model usage \cite{song2025modelgovernance}. We experimentally evaluated different LLM combinations to study optimal agent portfolios balancing cost-effectiveness, regulatory compliance, and optimization quality.

The aggregator LLM actively combines beneficial elements from multiple proposals while avoiding conflicting modifications, enabling effective synthesis even when limited to open-source models rather than using simple voting mechanisms.

\subsection{Key Design Decisions: Why MoA vs. GA Approaches}
Our decision to experiment with MoA alongside the company's vanilla GA stems from fundamental algorithmic differences that impact industrial applicability, particularly in regulated environments. 
Unlike the prompt-based optimization that GA employs, where improvements are achieved through iterative mutation and crossover of input prompts, MoA enables solution-based optimization by directly synthesizing code suggestions from multiple agents, potentially offering more comprehensive optimizations while accommodating regulatory constraints.

Also, MoA implements a feed-forward architecture with predictable execution patterns and fixed termination points, contrasting with GA's adaptive termination strategy where the stopping point varies based on convergence criteria, making execution time less predictable. MoA's fixed number of agent suggestions can reduce risks of single-point failures and bias while improving generalization, whereas GA's genome selection may identify superior solutions but risks introducing selection bias. These characteristics have significant implications for environments with strict operational requirements.


\section{Experimental Setup}



To evaluate the effectiveness of our MoA approach in industrial code optimization, particularly in environments with regulatory constraints, we designed a comprehensive experimental framework that assesses both optimization quality and operational efficiency across real-world codebases. Our experimental design addresses the following research questions:

\begin{itemize}
\item \textbf{RQ1:} How does model composition affect the performance of MoA and GA approaches in code optimization?
\item \textbf{RQ2:} How do MoA and GA approaches compare in terms of cost-effectiveness across different LLM combinations?
\item \textbf{RQ3:} What are the time-efficiency characteristics of MoA and GA approaches, particularly in regulated environments with model usage restrictions?
\end{itemize}

These research questions target critical concerns for industrial practitioners operating under regulatory constraints that limit model selection. Our framework simulates real-world regulatory scenarios by evaluating combinations ranging from commercial-only to open-source-only models, providing actionable insights through rigorous comparative evaluation.

\subsection{Dataset and Code Selection}

Our evaluation dataset consists of real-world code snippets extracted from the company's private software project on GitHub. This industrial codebase gives authentic optimization challenges reflecting the complexity and diversity of production environments, representing various programming constructs, algorithmic patterns, and optimization opportunities.

We manually selected 50 representative code snippets from a single directory spanning different complexity levels, functional domains, and optimization potential. This selection ensures comprehensive coverage of industrial optimization scenarios while enabling correct assessment of runtime impact from individual code modifications.

\subsection{Evaluation Methodology}

The evaluation systematically compares optimization outputs through a comprehensive experimental design across 50 selected code snippets using different LLM combinations.

\textbf{LLM-based Evaluation.} We employ GPT-4o-mini as an independent evaluator through ELO-rating-based LLM scoring for robust optimization assessment. The ELO rating system, originally developed for chess rankings \cite{elo1978rating}, performs pairwise comparisons of optimization variants to determine superior performance in head-to-head evaluations. For example, if there are 5 variants for an original code snippet, the system performs 10 pairwise comparisons, asking ``which is better''-based questions to output binary outcomes. We adopted this approach because single-pass LLM scoring can exhibit randomness and context sensitivity, while pairwise comparisons provide richer contextual evaluation by directly contrasting alternatives. These binary outcomes feed into standard ELO calculations, producing relative performance rankings across all optimization approaches while providing stronger validation and bias mitigation than one-time scoring methods \cite{chatbotarena_chiang2024}.

\textbf{Performance Analysis (RQ1).} We evaluate four combinations (COMB1-COMB4) selecting 3 out of 5 LLMs: llama-3-1-405b, vertex-claude-v35-sonnet, gemini-v25-flash, mistral-large-2, and gpt-o1-mini. This strategy enables verification of ensemble superiority by comparing GA and MoA performance against their constituent individual LLMs while maintaining computational feasibility. We specifically designed combinations that simulate environments with varying regulatory constraints, from fully commercial to primarily open-source. We perform per-combination ranking by evaluating ELO rates by model within each combination across five approaches: GA, MoA, and the three individual LLMs.

\textbf{Cost and Time Analysis (RQ2-3).} Three additional combinations (COMB5-COMB7) select 3 out of 10 LLMs, expanding to include deepseek-v3, gpt-o-3-mini, gpt-41-nano, and vertex-claude-v37-sonnet. This provides comprehensive cost-performance trade-off analysis across diverse model portfolios, where cost is quantified based on API usage charges including token consumption and model access fees. Cross-combination ranking focuses exclusively on GA and MoA results, identifying which optimization approach performs best under specific LLM combinations and revealing optimal configuration strategies for different industrial deployment scenarios.

Table \ref{tab:combinations} shows the seven LLM combinations used for experimental evaluation of GA and MoA, with COMB1-COMB4 used for performance analysis and COMB5-COMB7 used for cost and time analysis.
The aggregator component of MoA experiments with two distinct models: DeepSeek R1 (open-source) for COMB1-COMB4 and GPT-o1-mini (commercial) for COMB5-COMB7.
In each combination, GA, MoA, and individual LLMs generate optimized variants, producing at least 250 variants per combination. Five executions ensure statistical stability, yielding over 8,750 total optimized snippets across seven combinations.




\begin{table}[!t]
\centering
\caption{LLM Combinations for Evaluation of GA and MoA}
\begin{adjustbox}{width=\columnwidth}
\begin{tabular}{clll}
\toprule
\textbf{Combination} & \multicolumn{3}{c}{\textbf{Models}} \\
\midrule
COMB1 & Claude-v35-sonnet & Gemini-v25-flash & Llama-3-1-405b  \\
COMB2 & Claude-v35-sonnet & Mistral-large-2 & GPT-o1-mini \\
COMB3 & GPT-o1-mini & Mistral-large-2 & Llama-3-1-405b \\
COMB4 & Claude-v35-sonnet & Gemini-v25-flash & GPT-o1-mini \\
COMB5 & Claude-v37-sonnet & Mistral-large-2 & GPT-41-nano \\
COMB6 & Deepseek-v3 & Mistral-large-2 & Llama-3-1-405b \\
COMB7 & GPT-o3-mini & GPT-41-nano & Gemini-v25-pro \\
\bottomrule
\end{tabular}
\end{adjustbox}
\label{tab:combinations}
\end{table}

\section{Results and Analysis}
Our evaluation across seven LLM combinations and 50 code snippets reveals distinct performance traits between models, with particular implications for regulated environments.

\subsection{RQ1: Performance Comparison}
\begin{figure*}[!htb]
\centering
\includegraphics[width=1.0\textwidth]{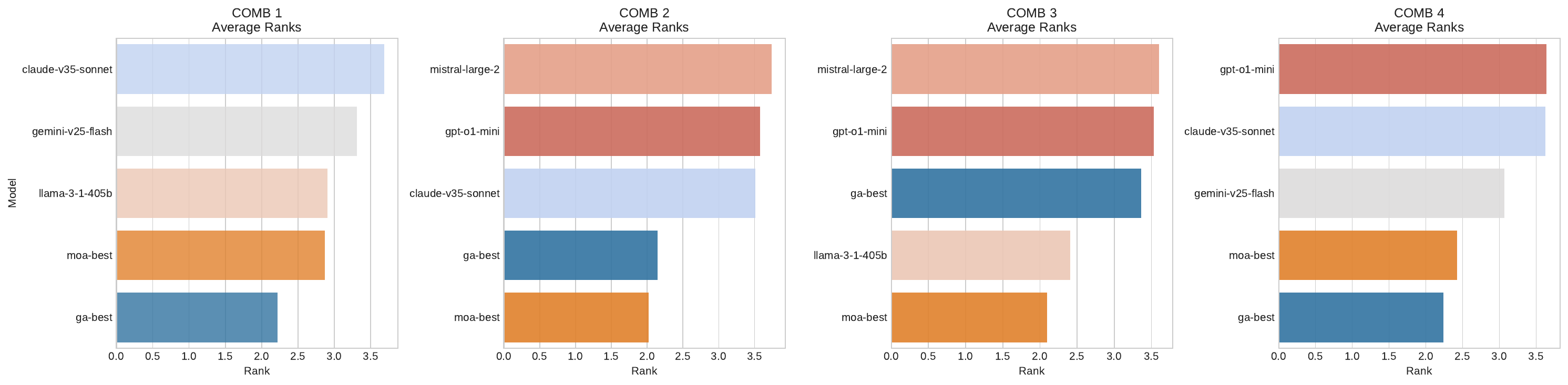}
\caption{Average ranks of ELO rates by combination. MoA excels with more open-source models (COMB3) and outperforms individual LLMs across all combinations; GA outperforms with commercial models (COMB4).}
\label{fig:bars_all}
\end{figure*}

Figure \ref{fig:bars_all} depicts average ranks of ELO rates by combination. MoA excels with more open-source models (COMB3) and outperforms individual LLMs across all combinations, while GA outperforms with commercial models (COMB4).
The performance analysis reveals distinct patterns based on model composition. For COMB4 (commercial models only), GA ranked best, demonstrating its strength when leveraging commercial models. MoA achieved optimal performance on COMB3 (2 open-source models and 1 commercial model), effectively leveraging diverse open-source optimization strategies where GA struggled, while consistently outperforming individual LLMs across all combinations.

This suggests that MoA's synthesis-based approach can be well-suited for combining outputs from heterogeneous open-source models, while GA's evolutionary approach excels when working with the consistent high-quality outputs of commercial models. These findings highlight the importance of matching optimization approach to model composition, with GA proving superior for commercial-heavy combinations and MoA demonstrating advantages in open-source-inclusive scenarios—particularly relevant for organizations operating under regulatory constraints that limit commercial model usage.

\subsection{RQ2: Cost-Effectiveness Analysis}
\begin{figure}[!htb]
\centering
\includegraphics[width=0.4\textwidth]{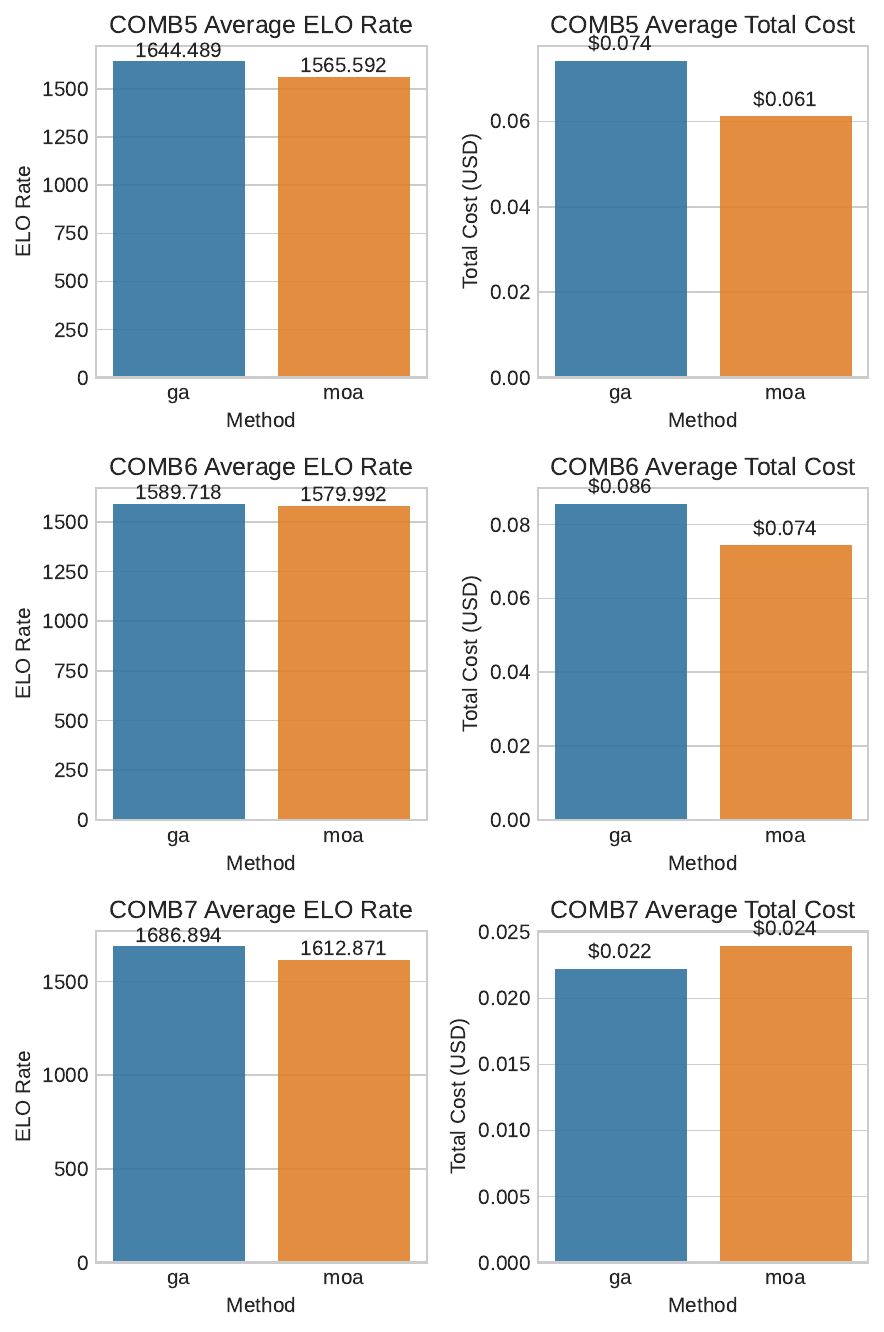}
\caption{Average ELO rate versus cost. MoA is 14.3\% to 22.2\% cheaper than GA for open-source combinations (COMB5-6), while GA is cheaper with commercial models (COMB7). Both achieve competitive ELO rates.}
\label{fig:bar}
\end{figure}

Figure \ref{fig:bar} presents cross-combination analysis using additional three LLM combinations, where ELO rates are computed across combinations to analyze cost-effectiveness variations by model composition. MoA saves cost 14.3\% to 22.2\% compared to GA for open-source combinations (COMB5-6), while GA is more economical with commercial models (COMB7).
This validates that MoA performs better with more open-source models, as shown in Figure \ref{fig:bars_all}. GA's advantage with COMB7 stems from its adaptive termination strategy—commercial models often achieve sufficient optimization in early generations, allowing GA to terminate with fewer recommendations and reduced costs, while MoA generates a fixed number of versions regardless of early convergence.

This cross-combination perspective enables identification of optimal deployment strategies based on available model portfolios and regulatory constraints, providing practical guidance for industrial cost-performance optimization with quantified savings for organizations using open-source models.

\subsection{RQ3: Time-Effectiveness Analysis}
\begin{figure}[!htb]
\centering
\includegraphics[width=0.4\textwidth]{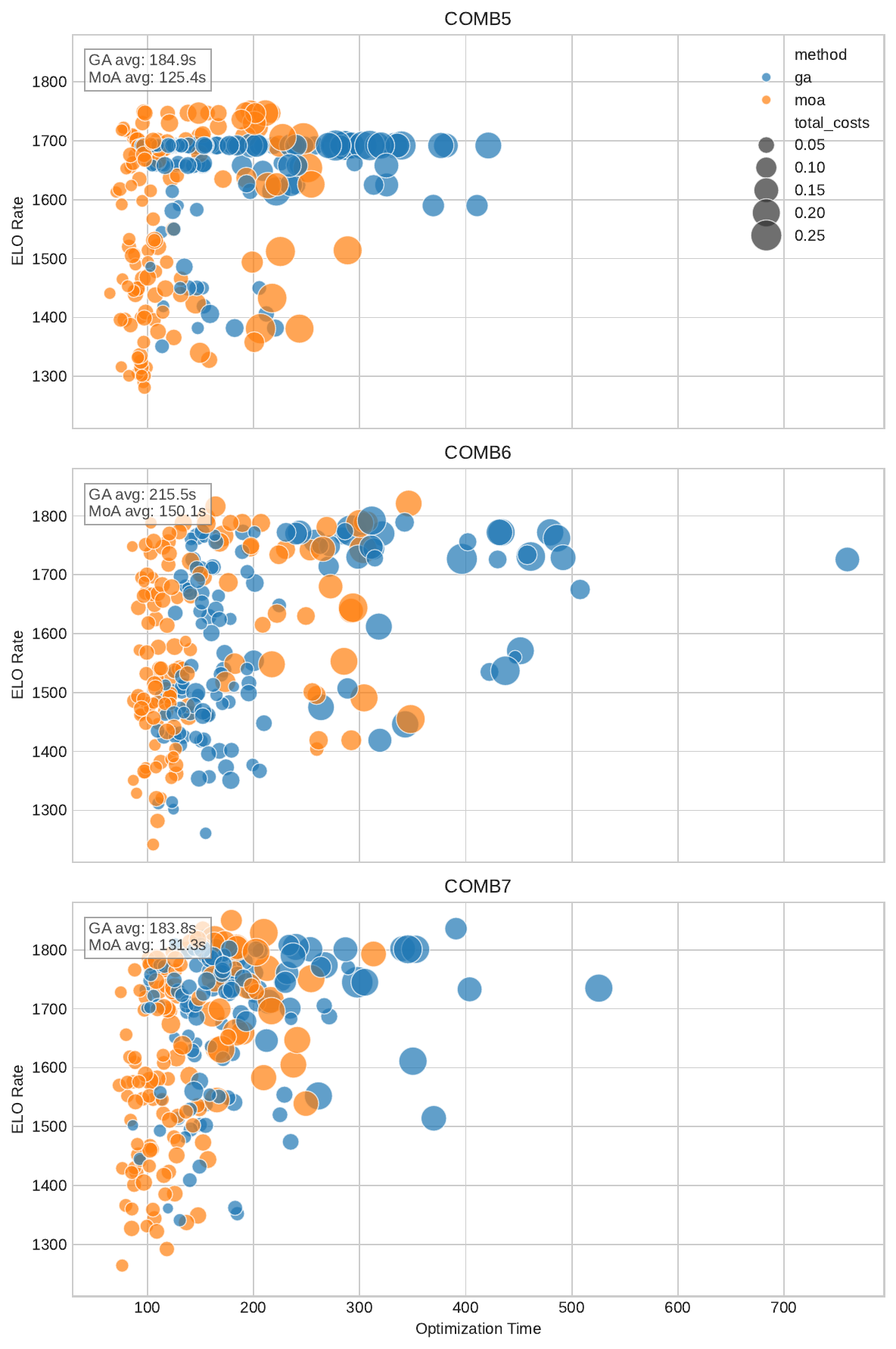}
\caption{ELO rate versus optimization time by combination. MoA consistently optimizes 28.6\% to 32.2\% faster than GA across combinations, while GA benefits from commercial models' rapid execution and high-quality outputs.}
\label{fig:scatter}
\end{figure}

Figure \ref{fig:scatter} shows ELO rate versus optimization time by combination, where dot size represents total cost per snippet. We filtered results to include only snippets with complete sets of six variants (2 approaches × 3 combinations from COMB5-7), excluding incomplete data due to processing failures. MoA consistently optimizes 28.6\% to 32.2\% faster than GA across combinations, providing significant advantages for time-sensitive deployments in regulated environments. However, GA benefits from commercial models' rapid execution and high-quality outputs, demonstrating notably faster performance for many snippets in COMB7.

The cost-performance relationship shows consistent patterns: longer optimization times correlate with higher costs (larger dots) for both approaches, indicating predictable cost-benefit trade-offs.

These findings reveal a clear model-composition dependency that provides actionable guidance for industrial deployment decisions based on available LLM resources and regulatory constraints.

\section{Discussion}

\textbf{Generalizability.} Our findings demonstrate high applicability of MoA on industrial code optimization through complementary deployment strategies: GA emerges as the preferred option when combining commercial models, leveraging their speed for cost-effective optimization, while MoA proves superior for combinations with open-source models, effectively synthesizing diverse optimization strategies with substantial time savings. Companies constrained by regulatory requirements (e.g., financial institutions prohibiting commercial models) can leverage MoA with open-source portfolios, while organizations with commercial model access can optimize costs through GA's adaptive termination strategy. 

\textbf{Scalability Considerations.} For larger codebases and development teams, our results suggest hybrid deployment strategies. MoA's consistent faster optimization times make it suitable for interactive development workflows where response time is critical. GA's cost-effectiveness with commercial models supports batch processing of large-scale optimization tasks. 
The predictable cost-benefit relationship enables budget-aware optimizations tailored to project and regulatory constraints. Even a \$0.01 per-snippet difference can quickly exceeds \$100 for large codebases with over 10,000 snippets, making the MoA vs. GA cost gap critical.

\textbf{Practical Recommendations}. Practitioners may match optimization approach to available model portfolios, e.g., deploy GA when using primarily commercial models, and MoA for open-source-inclusive combinations.
GA's struggles with open-source models stem from its indirect optimization approach—since our default GA-based ensemble system does not modify output code snippets directly, it has limited ability to overcome the inherent generation limitations of open-source models.
In contrast, MoA directly modifies and synthesizes output codes, effectively combining the best elements from multiple open-source model outputs to transcend individual model limitations. Organizations should implement these approaches with dynamic selection based on real-time cost estimates and time constraints.

\textbf{Future Directions}. 
Future work will involve improving and comparing our MoA approach against both existing ensemble methods and the company's advanced GA versions. This includes benchmarking against recent public agentic pipelines to compare routing vs. synthesis strategies, and implementing a cascading LLM pipeline to evaluate cost-optimization techniques. We also plan to measure actual execution time and CPU consumption to concretely validate performance improvements. The ELO-rating-based LLM scoring can also be improved with recent bias mitigation techniques \cite{elorating_liu2025}. Exploration of hybrid approaches combining GA's adaptive termination with MoA's synthesis strategies presents promising optimization opportunities.

\section{Conclusion}
This study addresses the challenge faced by organizations in regulated industries where compliance requirements restrict commercial LLM usage for code optimization. Our evaluation across 8,700+ optimized code snippets from real-world industrial codebases demonstrates that both MoA and GA ensemble approaches substantially improve over individual LLM optimizers, with distinct advantages in different regulatory contexts.

The key insight is model-composition dependency: MoA excels with open-source model combinations, providing a viable solution for regulated environments through effective synthesis of diverse optimization strategies with 14.3\% to 22.2\% cost savings and 28.6\% to 32.2\% faster optimization times. Conversely, GA performs optimally with commercial models through adaptive resource utilization when permitted.

These findings offer practical guidance for implementing ensemble-based code optimization while addressing broader industrial concerns of resource efficiency and runtime optimization.

\bibliography{bibtex/bib/IEEEabrv.bib,bibtex/bib/IEEEexample.bib}{}

\begin{thebibliography}{31}
\providecommand{\natexlab}[1]{#1}
\providecommand{\url}[1]{#1}
\csname url@samestyle\endcsname
\providecommand{\newblock}{\relax}
\providecommand{\bibinfo}[2]{#2}
\providecommand{\BIBentrySTDinterwordspacing}{\spaceskip=0pt\relax}
\providecommand{\BIBentryALTinterwordstretchfactor}{4}
\providecommand{\BIBentryALTinterwordspacing}{\spaceskip=\fontdimen2\font plus
\BIBentryALTinterwordstretchfactor\fontdimen3\font minus \fontdimen4\font\relax}
\providecommand{\BIBforeignlanguage}[2]{{%
\expandafter\ifx\csname l@#1\endcsname\relax
\typeout{** WARNING: IEEEtranN.bst: No hyphenation pattern has been}%
\typeout{** loaded for the language `#1'. Using the pattern for}%
\typeout{** the default language instead.}%
\else
\language=\csname l@#1\endcsname
\fi
#2}}
\providecommand{\BIBdecl}{\relax}
\BIBdecl

\bibitem[Giavrimis et~al.(2025)Giavrimis, Basios, Wu, Kanthan, and Bauer]{artemis_2025}
R.~Giavrimis, M.~Basios, F.~Wu, L.~Kanthan, and R.~Bauer, ``Artemis ai: Multi-llm framework for code optimisation,'' in \emph{2025 IEEE Conference on Artificial Intelligence (CAI)}, 2025, pp. 1--6.

\bibitem[Song(2025)]{song2025modelgovernance}
\BIBentryALTinterwordspacing
P.~Song, ``Model governance and compliance for regulated industries,'' ML Journey (blog), Jul. 14 2025, accessed: 2025-08-01. [Online]. Available: \url{https://mljourney.com/model-governance-and-compliance-for-regulated-industries/}
\BIBentrySTDinterwordspacing

\bibitem[Gong et~al.(2025)Gong, Voskanyan, Brookes, Wu, Jie, Xu, Giavrimis, Basios, Kanthan, and Wang]{survey_jing2025}
J.~Gong, V.~Voskanyan, P.~Brookes, F.~Wu, W.~Jie, J.~Xu, R.~Giavrimis, M.~Basios, L.~Kanthan, and Z.~Wang, ``Language models for code optimization: Survey, challenges and future directions,'' 2025, arXiv:2501.01277.

\bibitem[Ashiga et~al.(2025)Ashiga, Jie, Wu, Voskanyan, Dinmohammadi, Brookes, Gong, and Wang]{llm_ensemble_ashiga2025}
M.~Ashiga, W.~Jie, F.~Wu, V.~Voskanyan, F.~Dinmohammadi, P.~Brookes, J.~Gong, and Z.~Wang, ``Ensemble learning for large language models in text and code generation: A survey,'' 2025, arXiv:2503.13505.

\bibitem[Wang et~al.(2024)]{mixtureofagents_wang2024}
J.~Wang \emph{et~al.}, ``Mixture-of-agents enhances large language model capabilities,'' 2024, arXiv:2406.04692.

\bibitem[Yue et~al.(2024)]{llmcascades_yue2024}
M.~Yue \emph{et~al.}, ``Large language model cascades with mixture of thoughts representations for cost-efficient reasoning,'' in \emph{ICLR}, WIE, AT, 2024.

\bibitem[Akiba et~al.(2024)]{GAMoEs_akiba2024}
T.~Akiba \emph{et~al.}, ``Evolutionary optimization of model merging recipes,'' 2024, arXiv:2403.13187.

\bibitem[Holland(1975)]{GA_holland1975}
J.~H. Holland, \emph{Adaptation in Natural and Artificial Systems}.\hskip 1em plus 0.5em minus 0.4em\relax Ann Arbor, MI: University of Michigan Press, 1975.

\bibitem[Yu et~al.(2024)]{dare_yu2024}
L.~Yu \emph{et~al.}, ``Language models are super mario: Absorbing abilities from homologous models as a free lunch,'' in \emph{ICML}, WIE, AT, 2024.

\bibitem[Jiang et~al.(2023)]{llmblender_jiang2023}
D.~Jiang \emph{et~al.}, ``{LLM}-blender: Ensembling large language models with pairwise ranking and generative fusion,'' in \emph{ACL}, TO, CA, 2023, pp. 14\,165--14\,178.

\bibitem[Wan et~al.(2024{\natexlab{a}})]{knowledgefusionwan2024}
F.~Wan \emph{et~al.}, ``Knowledge fusion of large language models,'' in \emph{ICLR}, WIE, AT, 2024.

\bibitem[Jiang et~al.(2024)]{mixtralexperts_jiang2024}
A.~Q. Jiang \emph{et~al.}, ``Mixtral of experts,'' 2024, arXiv:2401.04088.

\bibitem[Yadav et~al.(2023)]{ties_merge_yadav2023}
P.~Yadav \emph{et~al.}, ``Ties-merging: Resolving interference when merging models,'' in \emph{NeurIPS}, vol.~36, LA, USA, 2023, pp. 7093--7115.

\bibitem[Wan et~al.(2024{\natexlab{b}})]{fusechat_wan2024}
F.~Wan \emph{et~al.}, ``Knowledge fusion of chat llms: A preliminary technical report,'' 2024, arXiv:2402.16107.

\bibitem[Duan et~al.(2025)Duan, Kanakaris, Xiao, Ping, Zhou, Ahmed, Ma, Capota, Willke, Nazarian, and Bogdan]{perfrl_duan2025}
S.~Duan, N.~Kanakaris, X.~Xiao, H.~Ping, C.~Zhou, N.~K. Ahmed, G.~Ma, M.~Capota, T.~L. Willke, S.~Nazarian, and P.~Bogdan, ``Perfrl: A small language model framework for efficient code optimization,'' 2025, arXiv:2312.05657.

\bibitem[Huang et~al.(2025)Huang, Dai, Weng, Wu, Qing, Cui, Guo, and Zhang]{effiLearner_huang2025}
D.~Huang, J.~Dai, H.~Weng, P.~Wu, Y.~Qing, H.~Cui, Z.~Guo, and J.~M. Zhang, ``Effilearner: Enhancing efficiency of generated code via self-optimization,'' 2025, arXiv:2405.15189.

\bibitem[Hong et~al.(2024)Hong, Bhatia, Haan, Dong, Nikiforov, Cheung, and Shao]{llmaided_hong2024}
C.~Hong, S.~Bhatia, A.~Haan, S.~K. Dong, D.~Nikiforov, A.~Cheung, and Y.~S. Shao, ``Llm-aided compilation for tensor accelerators,'' 2024, arXiv:2408.03408.

\bibitem[Nichols et~al.(2024)Nichols, Polasam, Menon, Marathe, Gamblin, and Bhatele]{performancealigned_nichols2024}
D.~Nichols, P.~Polasam, H.~Menon, A.~Marathe, T.~Gamblin, and A.~Bhatele, ``Performance-aligned llms for generating fast code,'' 2024, arXiv:2404.18864.

\bibitem[Romera-Paredes et~al.(2024)Romera-Paredes, Barekatain, Novikov, et~al.]{mathdiscovery_romera2024}
B.~Romera-Paredes, M.~Barekatain, A.~Novikov \emph{et~al.}, ``Mathematical discoveries from program search with large language models,'' \emph{Nature}, vol. 625, no. 7995, pp. 468--475, 2024.

\bibitem[Hemberg et~al.(2024)Hemberg, Moskal, and O’Reilly]{evolcode_hemberg2024}
E.~Hemberg, S.~Moskal, and U.-M. O’Reilly, ``Evolving code with a large language model,'' \emph{Genetic Programming and Evolvable Machines}, vol.~25, no.~2, Sep. 2024.

\bibitem[Ilharco et~al.(2023)]{taskarithmetic}
G.~Ilharco \emph{et~al.}, ``Editing models with task arithmetic,'' in \emph{ICLR}, 2023.

\bibitem[Fedus et~al.(2022)]{switchtransformer_fedus2022}
W.~Fedus \emph{et~al.}, ``Switch transformers: Scaling to trillion parameter models with simple and efficient sparsity,'' in \emph{JMLR}, vol.~23, no. 120, 2022, pp. 1--39.

\bibitem[Coste et~al.(2024)]{rewardmodelensembles_coste2024}
T.~Coste \emph{et~al.}, ``Reward model ensembles help mitigate overoptimization,'' in \emph{ICLR}, WIE, AT, 2024.

\bibitem[Zhang et~al.(2024)]{rewardmodelensembleRlhf_zhang2024}
S.~Zhang \emph{et~al.}, ``Improving reinforcement learning from human feedback with efficient reward model ensemble,'' 2024, arXiv:2401.16635.

\bibitem[Mohammadshahi et~al.(2024)]{routoo_mohammadshahi2024}
A.~Mohammadshahi \emph{et~al.}, ``Routoo: Learning to route to large language models effectively,'' 2024, arXiv:2401.13979.

\bibitem[Chen et~al.(2023)]{frugalgpt_chen2023}
L.~Chen \emph{et~al.}, ``Frugalgpt: How to use large language models while reducing cost and improving performance,'' 2023, arXiv:2305.05176.

\bibitem[Lab(2023)]{alpacaeval2023}
\BIBentryALTinterwordspacing
T.~Lab, ``Alpacaeval: An automatic evaluator for instruction-following language models,'' 2023, accessed: 2025-06-06. [Online]. Available: \url{https://github.com/tatsu-lab/alpaca_eval}
\BIBentrySTDinterwordspacing

\bibitem[Zheng et~al.(2023)]{mt-bench_zheng2023}
L.~Zheng \emph{et~al.}, ``Judging llm-as-a-judge with mt-bench and chatbot arena,'' in \emph{NeurIPS}, 2023.

\bibitem[Elo(1978)]{elo1978rating}
A.~Elo, \emph{The Rating of Chessplayers, Past and Present}.\hskip 1em plus 0.5em minus 0.4em\relax Arco Pub., 1978.

\bibitem[Chiang et~al.(2024)Chiang, Zheng, Sheng, Angelopoulos, Li, Li, Zhu, Zhang, Jordan, Gonzalez, and Stoica]{chatbotarena_chiang2024}
W.-L. Chiang, L.~Zheng, Y.~Sheng, A.~N. Angelopoulos, T.~Li, D.~Li, B.~Zhu, H.~Zhang, M.~I. Jordan, J.~E. Gonzalez, and I.~Stoica, ``Chatbot arena: an open platform for evaluating llms by human preference,'' in \emph{Proceedings of the 41st International Conference on Machine Learning}, ser. ICML'24.\hskip 1em plus 0.5em minus 0.4em\relax JMLR.org, 2024.

\bibitem[Liu et~al.(2025)Liu, Gemp, Marris, Piliouras, Heess, and Lanctot]{elorating_liu2025}
S.~Liu, I.~Gemp, L.~Marris, G.~Piliouras, N.~Heess, and M.~Lanctot, ``Re-evaluating open-ended evaluation of large language models,'' in \emph{ICLR}, 2025.

\end{thebibliography}
\bibliographystyle{IEEEtranN}

\end{document}